\documentclass[
a4paper,%
aps,%
12pt,%
final,%
notitlepage,%
oneside,%
onecolumn,%
nobibnotes,%
nofootinbib,%
superscriptaddress,%
noshowpacs]%
{revtex4}
\usepackage{graphicx}
\usepackage{color}
\usepackage{amsmath}
\usepackage{amsfonts}
\usepackage{amssymb}
\usepackage{accents}
\usepackage{mathtools}
\usepackage[utf8]{inputenc}
\usepackage[T1]{fontenc}
\usepackage[colorlinks=true]{hyperref}

\newcommand{\order}[2]{\accentset{#2}{#1}}
\newcommand{\gauge}[2]{\prescript{\mathcal{#2}}{}{#1}}
\newcommand{\gaor}[3]{\prescript{\mathcal{#2}}{}{\accentset{#3}{#1}}}

\newcommand{\dd}{\mathrm{d}}

\begin{document}

\noindent {\it Astronomy Reports, 2021, Vol. 65, No. 10}
\bigskip\bigskip  \hrule\smallskip\hrule
\vspace{35mm}


\title{Gauge-invariant post-Newtonian perturbations in symmetric teleparallel gravity\footnote{Paper presented at the Fourth Zeldovich meeting, an international conference in honor of Ya. B. Zeldovich held in Minsk, Belarus on September 7--11, 2020. Published by the recommendation of the special editors: S. Ya. Kilin, R. Ruffini and G. V. Vereshchagin.}}

\author{\bf \copyright $\:$  2021.
\quad \firstname{Manuel}~\surname{Hohmann}}%
\email{manuel.hohmann@ut.ee}
\affiliation{Laboratory of Theoretical Physics, Institute of Physics, University of Tartu, W. Ostwaldi 1, 50411 Tartu, Estonia}%

\begin{abstract}
\centerline{\footnotesize Received: ;$\;$
Revised: ;$\;$ Accepted: .}\bigskip\bigskip\bigskip

We present an extension to the gauge-invariant formulation of the parameterized post-Newtonian (PPN) formalism, which allows its application to symmetric teleparallel gravity theories. In its original formulation, the gauge-invariant PPN formalism makes use of a gauge-invariant perturbative expansion of tensor fields; however, one of the fundamental gravitational field variables in symmetric teleparallel gravity theories is a flat, torsion-free connection, hence not a tensor field. Since connections transform differently from tensor fields under diffeomorphisms, we introduce here an adapted notion of gauge-invariant perturbation variables for the symmetric teleparallel connection, and show how the lowest order terms can be expressed in terms of the common PPN potentials.
\end{abstract}

\maketitle

\section{Introduction}
The open questions in cosmology and the tensions between general relativity and quantum theory have given rise to the study of various modified gravity theories. While most of these modifications depart from the formulation of general relativity in terms of the curvature of the Levi-Civita connection of the spacetime metric, alternative formulations in terms of the torsion or nonmetricity of a flat connection have received recent interest~\cite{BeltranJimenez:2019tjy}. Here we focus on theories which arise as modifications of the symmetric teleparallel equivalent of general relativity (STEGR)~\cite{Nester:1998mp}, and which employ the nonmetricity of a flat, torsion-free connection as the mediator of gravity. Notable contender theories include newer general relativity~\cite{BeltranJimenez:2017tkd,BeltranJimenez:2018vdo}, $f(Q)$ theories~\cite{BeltranJimenez:2017tkd,BeltranJimenez:2018vdo}, scalar-nonmetricity gravity~\cite{Jarv:2018bgs,Runkla:2018xrv} and generalizations thereof~\cite{Hohmann:2018xnb,Koivisto:2019jra}.

An important tool for testing the validity of such modified gravity theories is the parametrized post-Newtonian (PPN) formalism~\cite{Will:1993ns,Will:2014kxa,Will:2018bme}. Its core technique is the perturbative expansion of the metric tensor, and possibly other fields, around a vacuum solution to the gravitational field equations, which is usually performed in a particularly chosen reference frame. In practice, this may lead to difficulties, since the chosen reference frame is fully determined only after solving the field equations. In order to address this issue, a gauge-invariant approach to the PPN formalism has been developed~\cite{Hohmann:2019qgo}, based on the theory of nonlinear gauge transformations~\cite{Bruni:1996im,Bruni:1999et}, which allows for the definition of a higher order Taylor expansion of tensor fields~\cite{Sonego:1997np}. This underlying mathematical foundation has formerly proven to be useful in its application to perturbations in more than one variable~\cite{Nakamura:2003wk,Nakamura:2004wr} as well as to higher order gauge-invariant cosmological perturbation theory~\cite{Nakamura:2004rm,Nakamura:2006rk}.

In the gauge-invariant PPN formalism, the fundamental fields which mediate the gravitational interaction are assumed to be a metric or a tetrad, which is understood as a Lorentz-valued covector field, and possibly other tensor fields; an example with a scalar field has been worked out~\cite{Hohmann:2019qgo}. In symmetric teleparallel gravity, however, as well as Palatini or metric-affine theories of gravity, a connection appears in the role of a gravitational field variable. Since connections transform differently from tensor fields under diffeomorphisms, which lie at the heart of the gauge-invariant perturbation theory, the question arises how these may be incorporated into the gauge-invariant PPN formalism. The aim of this article is to demonstrate this in case of the flat, torsion-free connection present in symmetric teleparallel gravity theories.

\section{Preliminaries}\label{sec:prelim}
Before presenting the main construction in this article, we briefly review the necessary preliminaries. The gauge-invariant PPN formalism is displayed in section~\ref{ssec:ginvppn}, while a short introduction to symmetric teleparallel gravity is given in section~\ref{ssec:symtele}.

\subsection{Gauge-invariant PPN formalism}\label{ssec:ginvppn}
We start with a brief review of the essential ingredients of the gauge-invariant PPN formalism; see~\cite{Hohmann:2019qgo} for a full discussion. The starting point of this formalism is the idea that physical fields, such as the metric \(g_{\mu\nu}\), are defined on a physical spacetime manifold \(M\), which is not equipped with any coordinates or background fields. The latter are defined only on a model manifold \(M_0\), which is equipped with a background metric \(g^{(0)}_{\mu\nu}\). In order to relate these different manifolds and tensors fields defined on them, one must make use of a \emph{gauge}, i.e., a diffeomorphism \(\mathcal{X}: M_0 \to M\), which allows to construct the pullback \(\gauge{g}{X}_{\mu\nu} = (\mathcal{X}^*g)_{\mu\nu}\) of the metric and other tensor fields to the background manifold \(M_0\). This pullback \(\gauge{g}{X}_{\mu\nu}\) is then regarded as a perturbation of the background metric \(g^{(0)}_{\mu\nu}\). In the case of the PPN formalism, this background manifold \(M_0\) is assumed to be Minkowski space, equipped with the background metric \(g^{(0)}_{\mu\nu} = \eta_{\mu\nu}\) and the usual Cartesian coordinates \((x^{\mu}) = (t, x^i)\).

Besides the pullback of tensor fields from \(M\) to \(M_0\), a gauge \(\mathcal{X}\) also induces coordinates \((\gauge{x}{X}^{\mu} = x^{\mu} \circ \mathcal{X}^{-1})\) on \(M\). This allows for two different viewpoints of gauge transformations, i.e., changes from one gauge \(\mathcal{X}\) to another gauge \(\mathcal{Y}\): actively, as a change of the pullback tensor fields \(\gauge{g}{X}_{\mu\nu}\) to \(\gauge{g}{Y}_{\mu\nu}\) at the same point with fixed coordinates \((x^{\mu})\) on \(M_0\), or passively, as expressing the same tensor field \(g_{\mu\nu}\) on \(M\) in different coordinates \((\gauge{x}{X}^{\mu})\) and \((\gauge{x}{Y}^{\mu})\). 

Perturbations of the metric and other tensor fields in a given gauge are defined as
\begin{equation}
\gauge{g}{X}_{\mu\nu} = \eta_{\mu\nu} + \gaor{g}{X}{1}_{\mu\nu} + \gaor{g}{X}{2}_{\mu\nu} + \gaor{g}{X}{3}_{\mu\nu} + \gaor{g}{X}{4}_{\mu\nu} + \mathcal{O}(5)\,,
\end{equation}
up to the fourth order, where the order parameter is the velocity of the source matter, and only certain components of these perturbations are relevant and non-vanishing; see~\cite{Will:1993ns,Will:2014kxa,Will:2018bme} for a full exposition. These perturbation components are then expressed in terms of Poisson-like integrals over the source matter variables (called the PPN potentials), with constant coefficients (called the PPN parameters). The latter depend on the gravity theory under consideration, and can be used to confront this theory with observations.

In the classical PPN formalism, a standard gauge \(\mathcal{P}\) is used, which is defined by the absence of certain PPN potentials from the perturbations \(\gaor{g}{P}{2}_{ij}\) and \(\gaor{g}{P}{4}_{00}\). The gauge-invariant PPN formalism, however, makes use of a different gauge \(\mathcal{S}\), which is defined by the purely geometric conditions that \(\gaor{g}{S}{2}_{ij}\) is diagonal, while \(\gaor{g}{S}{3}_{0i}\) is divergence-free. Decomposing the metric perturbation into irreducible components then implies that its non-vanishing components are gauge-invariant. These gauge-invariant components are denoted with boldface letters, \(\order{\mathbf{g}}{k}_{\mu\nu} \equiv \gaor{g}{S}{k}_{\mu\nu}\). We will employ the same convention here.

\subsection{Symmetric teleparallel gravity}\label{ssec:symtele}
In symmetric teleparallel gravity, one considers a connection with coefficients \(\Gamma^{\mu}{}_{\nu\rho}\) as dynamical gravitational field variable, in addition to the metric \(g_{\mu\nu}\). This connection is assumed to be flat and torsion-free, hence
\begin{equation}
R^{\rho}{}_{\sigma\mu\nu} = \partial_{\mu}\Gamma^{\rho}{}_{\sigma\nu} - \partial_{\nu}\Gamma^{\rho}{}_{\sigma\mu} + \Gamma^{\rho}{}_{\tau\mu}\Gamma^{\tau}{}_{\sigma\nu} - \Gamma^{\rho}{}_{\tau\nu}\Gamma^{\tau}{}_{\sigma\mu} \equiv 0\,, \quad
T^{\rho}{}_{\mu\nu} = \Gamma^{\rho}{}_{\nu\mu} - \Gamma^{\rho}{}_{\mu\nu} \equiv 0\,.
\end{equation}
The dynamics for the fundamental field variables are encoded in the nonmetricity, or equivalently the disformation tensors, defined by
\begin{equation}\label{eq:affnonmet}
Q_{\rho\mu\nu} = \nabla_{\rho}g_{\mu\nu}\,, \quad
L^{\mu}{}_{\nu\rho} = \frac{1}{2}\left(Q^{\mu}{}_{\nu\rho} - Q_{\nu}{}^{\mu}{}_{\rho} - Q_{\rho}{}^{\mu}{}_{\nu}\right)\,,
\end{equation}
which are in general non-vanishing. In terms of the latter, the action for STEGR reads
\begin{equation}\label{eq:stegraction}
S_{\text{STEGR}} = \frac{1}{2\kappa^2}\int\dd^4x\,\sqrt{-g}(L^{\mu\nu\rho}L_{\nu\mu\rho} - L^{\mu}{}_{\mu\rho}L^{\rho\nu}{}_{\nu})\,,
\end{equation}
with gravitational constant \(\kappa^2 = 8\pi G\). Further, one conventionally assumes that there is no direct coupling between the symmetric teleparallel connection and matter fields, so that the gravitational action is supplemented with a matter action \(S_{\text{m}}[g, \chi]\) depending only on the metric \(g_{\mu\nu}\) and some arbitrary matter fields \(\chi\). While it is possible to relax this assumption~\cite{BeltranJimenez:2020sih}, here we choose to retain it, as it guarantees that the trajectories of test bodies will be given by the geodesics of the metric, which is an important assumption of the PPN formalism, and provides the link between the PPN parameters and observations~\cite{Will:1993ns,Will:2014kxa,Will:2018bme}.

\section{Extension of the formalism}\label{sec:extension}
We now extend the gauge-invariant PPN formalism to the case of symmetric teleparallel gravity theories. First, we discuss the gauge transformation of connections in section~\ref{ssec:conngatra}. We then construct gauge-invariant perturbation variables for the symmetric teleparallel connection in section~\ref{ssec:stppert}, which we then express in PPN potentials in section~\ref{ssec:ppnpot}.

\subsection{Gauge transformation of connections}\label{ssec:conngatra}
The original gauge-invariant PPN formalism~\cite{Hohmann:2019qgo} is based on the theory of gauge transformations of tensor fields~\cite{Bruni:1996im,Bruni:1999et}. The coefficients of a connection, however, do not form a tensor field. Their gauge transformations can most easily be understood from the passive interpretation of diffeomorphisms as coordinate changes, from which follows that a connection transforms by the well-known formula
\begin{equation}\label{eq:connfintra}
\gauge{\Gamma}{Y}^{\mu}{}_{\nu\rho} = \gauge{\Gamma}{X}^{\alpha}{}_{\beta\gamma}\frac{\partial\gauge{x}{Y}^{\mu}}{\partial\gauge{x}{X}^{\alpha}}\frac{\partial\gauge{x}{X}^{\beta}}{\partial\gauge{x}{Y}^{\nu}}\frac{\partial\gauge{x}{X}^{\gamma}}{\partial\gauge{x}{Y}^{\rho}} + \frac{\partial\gauge{x}{Y}^{\mu}}{\partial\gauge{x}{X}^{\sigma}}\frac{\partial^2\gauge{x}{X}^{\sigma}}{\partial\gauge{x}{Y}^{\nu}\partial\gauge{x}{Y}^{\rho}}\,.
\end{equation}
In the perturbative approach, the gauge transformation is parametrized by a sequence of vector fields \(\order{\xi}{1}^{\mu}, \order{\xi}{2}^{\mu}, \ldots\) on \(M_0\), which at the lowest orders can be written as
\begin{equation}\label{eq:coordtrans}
\gauge{x}{X}^{\mu} = \gauge{x}{Y}^{\mu} + \order{\xi}{1}^{\mu} + \order{\xi}{2}^{\mu} + \frac{1}{2}\order{\xi}{1}^{\nu}\partial_{\nu}\order{\xi}{1}^{\mu} + \mathcal{O}(3)\,.
\end{equation}
Performing a perturbative expansion of the finite coordinate transformation~\eqref{eq:connfintra} in the vector fields \(\xi\), and considering the connection to be given as a perturbation around a given background \(\order{\Gamma}{0}^{\mu}{}_{\nu\rho}\) on \(M_0\), one finds at the lowest orders the terms
\begin{gather}
\gaor{\Gamma}{Y}{0}^{\mu}{}_{\nu\rho} = \gaor{\Gamma}{X}{0}^{\mu}{}_{\nu\rho} = \order{\Gamma}{0}^{\mu}{}_{\nu\rho}\,, \quad
\gaor{\Gamma}{Y}{1}^{\mu}{}_{\nu\rho} = \gaor{\Gamma}{X}{1}^{\mu}{}_{\nu\rho} + (\mathcal{L}_{\order{\xi}{1}}\gaor{\Gamma}{X}{0})^{\mu}{}_{\nu\rho}\,,\nonumber\\
\gaor{\Gamma}{Y}{2}^{\mu}{}_{\nu\rho} = \gaor{\Gamma}{X}{2}^{\mu}{}_{\nu\rho} + (\mathcal{L}_{\order{\xi}{1}}\gaor{\Gamma}{X}{1})^{\mu}{}_{\nu\rho} + (\mathcal{L}_{\order{\xi}{2}}\gaor{\Gamma}{X}{0})^{\mu}{}_{\nu\rho} + \frac{1}{2}(\mathcal{L}_{\order{\xi}{1}}\mathcal{L}_{\order{\xi}{1}}\gaor{\Gamma}{X}{0})^{\mu}{}_{\nu\rho}\,,\label{eq:traconn}
\end{gather}
which are formally identical to the gauge transformation of tensor components, with the only exception arising from the fact that the Lie derivative of a connection receives an additional contribution from the inhomogeneous term in the transformation~\eqref{eq:connfintra} and reads~\cite{Yano:1957lda}
\begin{equation}\label{eq:connlieder}
(\mathcal{L}_{\xi}\Gamma)^{\mu}{}_{\nu\rho} = \xi^{\sigma}\partial_{\sigma}\Gamma^{\mu}{}_{\nu\rho} - \partial_{\sigma}\xi^{\mu}\Gamma^{\sigma}{}_{\nu\rho} + \partial_{\nu}\xi^{\sigma}\Gamma^{\mu}{}_{\sigma\rho} + \partial_{\rho}\xi^{\sigma}\Gamma^{\mu}{}_{\nu\sigma} + \partial_{\nu}\partial_{\rho}\xi^{\mu}\,.
\end{equation}
This Lie derivative~\eqref{eq:connlieder}, however, is again a tensor field, and so any higher order Lie derivatives are simply obtained by the usual homogeneous Lie derivative formula.

\subsection{Perturbative expansion of the symmetric teleparallel connection}\label{ssec:stppert}
An important property of the symmetric teleparallel connection, which follows from the conditions that it is flat and symmetric, is the existence of a gauge \(\mathcal{C}\), called the \emph{coincident gauge}~\cite{BeltranJimenez:2017tkd}, in which its connection coefficients vanish identically, \(\gauge{\Gamma}{C}^{\mu}{}_{\nu\rho} \equiv 0\). Note that this condition does not determine the coincident gauge uniquely; any other gauge \(\tilde{\mathcal{C}}\) which is related by a linear coordinate transformation
\begin{equation}\label{eq:lincoord}
\prescript{\tilde{\mathcal{C}}}{}{x}^{\mu} = \gauge{x}{C}^{\mu} + x_0^{\mu} + \Lambda^{\mu}{}_{\nu}\gauge{x}{C}^{\nu}
\end{equation}
with constant \(x_0^{\mu}\) and \(\Lambda^{\mu}{}_{\nu}\) also satisfies \(\prescript{\tilde{\mathcal{C}}}{}{\Gamma}^{\mu}{}_{\nu\rho} \equiv 0\)~\cite{BeltranJimenez:2018vdo}. Further, in general, none of these coincident gauges agrees with either the distinguished gauge \(\mathcal{S}\) with respect to which the gauge-invariant metric perturbations are defined, nor with the standard PPN gauge \(\mathcal{P}\). However, we will make the assumption that there exists a coincident gauge \(\mathcal{C}\) which is related to these gauges by a gauge transformation which preserves the PPN order of perturbations. This assumption is justified by its implication that the resulting nonmetricity and disformation tensors, which enter the gravitational field equations, will then be of PPN order, with a common vacuum solution given by \(\order{g}{0}_{\mu\nu} = \eta_{\mu\nu}\) and \(\order{\Gamma}{0}^{\mu}{}_{\nu\rho} = 0\), so that the gravitational side of the field equations becomes of the same order as the energy-momentum side. It thus follows that the vector field \(C\), which relates the coincident gauge \(\mathcal{C}\) to the distinguished gauge \(\mathcal{S}\), is described by the only relevant and non-vanishing components
\begin{equation}\label{eq:coincomp}
\order{C}{2}^i\,, \quad
\order{C}{3}^0\,, \quad
\order{C}{4}^i\,.
\end{equation}
From the defining property \(\gauge{\Gamma}{C}^{\mu}{}_{\nu\rho} \equiv 0\) of the coincident gauge, as well as the gauge transformation~\eqref{eq:traconn} with \(\mathcal{Y} = \mathcal{C}\), \(\mathcal{X} = \mathcal{S}\) and \(\xi = C\) (or, equivalently, \(\mathcal{Y} = \mathcal{S}\), \(\mathcal{X} = \mathcal{C}\) and \(\xi = -C\)), then follows that in the distinguished gauge the symmetric teleparallel connection reads
\begin{gather}
\gaor{\Gamma}{S}{2}^i{}_{jk} = -\partial_j\partial_k\order{C}{2}^i\,, \quad
\gaor{\Gamma}{S}{3}^i{}_{j0} = -\partial_j\partial_0\order{C}{2}^i\,, \quad
\gaor{\Gamma}{S}{3}^0{}_{jk} = -\partial_j\partial_k\order{C}{3}^0\,, \quad
\gaor{\Gamma}{S}{4}^i{}_{00} = -\partial_0\partial_0\order{C}{2}^i\,,\label{eq:symteleconn}\\
\gaor{\Gamma}{S}{4}^0{}_{j0} = -\partial_j\partial_0\order{C}{3}^0\,, \quad
\gaor{\Gamma}{S}{4}^i{}_{jk} = -\partial_j\partial_k\order{C}{4}^i + \frac{1}{2}\left(\order{C}{2}^l\partial_j\partial_k\partial_l\order{C}{2}^i + 2\partial_{(j}\order{C}{2}^l\partial_{k)}\partial_l\order{C}{2}^i - \partial_j\partial_k\order{C}{2}^l\partial_l\order{C}{2}^i\right)\,,\nonumber
\end{gather}
which can also easily be derived from the finite transformation law~\eqref{eq:connfintra} for the connection coefficients and series expansion of the coordinate transformation~\eqref{eq:coordtrans}. We see from the appearance of derivatives of second order in the relations~\eqref{eq:symteleconn} that linear coordinate transformations of the form~\eqref{eq:lincoord} indeed constitute a freedom in the choice of the coincident gauge. However, this residual gauge freedom is fixed by the demand that the gauge transforming vector field \(C^{\mu}\) is of PPN order~\eqref{eq:coincomp}, which does not allow for constant or linear contributions. Hence, this condition uniquely fixes a coincident gauge. The components~\eqref{eq:coincomp} of the gauge defining vector field thus constitute the perturbative expansion of the gauge-invariant symmetric teleparallel connection coefficients \(\boldsymbol{\Gamma}^{\mu}{}_{\nu\rho} \equiv \gauge{\Gamma}{S}^{\mu}{}_{\nu\rho}\), in analogy to the gauge invariant metric components \(\mathbf{g}_{\mu\nu}\) introduced in section~\ref{ssec:ginvppn}, and we can indicate this by promoting the notation to use boldface letters \(\mathbf{C}^{\mu}\).

\subsection{Expansion in PPN potentials}\label{ssec:ppnpot}
We finally come to the question how to express the relevant components~\eqref{eq:coincomp} of the gauge defining vector field \(C^{\mu}\), which constitute the coefficients \(\gauge{\Gamma}{S}^{\mu}{}_{\nu\rho}\) of the symmetric teleparallel connection in the distinguished gauge \(\mathcal{S}\), in terms of the PPN potentials. For this purpose we recall that the non-vanishing components of the energy-momentum tensor in the same gauge, up to the relevant perturbation orders, are given by~\cite{Hohmann:2019qgo}
\begin{equation}\label{eq:enmomppn}
\order{\mathbf{T}}{2}_{00} = \boldsymbol{\rho}\,, \quad
\order{\mathbf{T}}{4}_{00} = \boldsymbol{\rho}\left(\mathbf{v}^2 + \boldsymbol{\Pi} - \order{\mathbf{g}}{2}_{00}\right)\,, \quad
\order{\mathbf{T}}{3}_{0i} = -\boldsymbol{\rho}\mathbf{v}_i\,, \quad
\order{\mathbf{T}}{4}_{ij} = \boldsymbol{\rho}\mathbf{v}_i\mathbf{v}_j + \mathbf{p}\delta_{ij}\,.
\end{equation}
This must be compared with the corresponding components of the gravitational side of the field equations of a given symmetric teleparallel gravity theory under consideration. Following section~\ref{ssec:symtele}, these are constituted by the nonmetricity tensor and its covariant derivatives, as well as possibly other tensor fields which are present in the gravitational action. Here we aim for a generic discussion, and consider only the most simple cases. Expanding the nonmetricity tensor based on its definition~\eqref{eq:affnonmet} into gauge-invariant perturbations of the metric and symmetric teleparallel connection yields the non-vanishing components
\begin{equation}
\order{\mathbf{Q}}{2}_{i00} = \partial_i\order{\mathbf{g}}{2}_{00}\,, \quad
\order{\mathbf{Q}}{2}_{ijk} = \partial_i\order{\mathbf{g}}{2}_{jk} + 2\partial_i\partial_{(j}\order{\mathbf{C}}{2}_{k)}
\end{equation}
at the second velocity order, as well as
\begin{equation}
\order{\mathbf{Q}}{3}_{000} = \partial_0\order{\mathbf{g}}{2}_{00}\,, \quad
\order{\mathbf{Q}}{3}_{0jk} = \partial_0\order{\mathbf{g}}{2}_{jk} + 2\partial_0\partial_{(j}\order{\mathbf{C}}{2}_{k)}\,, \quad
\order{\mathbf{Q}}{3}_{ij0} = \partial_i\order{\mathbf{g}}{3}_{j0} - \partial_i\partial_j\order{\mathbf{C}}{3}_0 + \partial_0\partial_i\order{\mathbf{C}}{2}_j
\end{equation}
at the third velocity order, where the metric perturbations are conventionally assumed to be of the form
\begin{equation}\label{eq:metricppn}
\order{\mathbf{g}}{2}_{00} = 2\mathbf{U}\,, \quad
\order{\mathbf{g}}{2}_{ij} = 2\gamma\mathbf{U}\delta_{ij}\,, \quad
\order{\mathbf{g}}{3}_{0i} = -\left(1 + \gamma + \frac{\alpha_1}{4}\right)(\mathbf{V}_i + \mathbf{W}_i)\,.
\end{equation}
Here \(\gamma\) and \(\alpha_1\) are (usually constant) PPN parameters, which depend on the chosen theory, and the PPN potentials can be obtained as the solutions to the Poisson-like equations
\begin{equation}\label{eq:ppnpotdef}
\triangle\boldsymbol{\chi} = -2\mathbf{U}\,, \quad
\triangle\mathbf{U} = -4\pi\boldsymbol{\rho}\,, \quad
\triangle\mathbf{V}_i = -4\pi\boldsymbol{\rho}\mathbf{v}_i\,, \quad
\triangle\mathbf{W}_i = -4\pi\boldsymbol{\rho}\mathbf{v}_i + 2\mathbf{U}_{,0i}\,.
\end{equation}
Relevant terms for the metric field equations of symmetric teleparallel gravity theories, which are symmetric rank-two tensor equations by construction, can be obtained either by considering products of nonmetricity terms, which are then at least of fourth velocity order, or derivatives of the nonmetricity. At the lowest orders, only the latter appear. Calculating all terms of this type, one finds at the second velocity order the terms
\begin{equation}
\partial_i\order{\mathbf{Q}}{2}_{i00}\,, \quad
\partial_i\order{\mathbf{Q}}{2}_{ijj}\,, \quad
\partial_i\order{\mathbf{Q}}{2}_{jij}\,, \quad
\partial_i\order{\mathbf{Q}}{2}_{j00}\,, \quad
\partial_i\order{\mathbf{Q}}{2}_{jkk}\,, \quad
\partial_i\order{\mathbf{Q}}{2}_{kjk}\,, \quad
\partial_k\order{\mathbf{Q}}{2}_{kij}\,, \quad
\partial_k\order{\mathbf{Q}}{2}_{(ij)k}\,.
\end{equation}
which are three scalars and five symmetric spatial two-tensors. Together with the expansion~\eqref{eq:enmomppn} of the energy-momentum tensor and~\eqref{eq:metricppn} of the metric, one finds that the field equations at the second velocity order can be brought into the generic form
\begin{equation}\label{eq:ppnord2}
a_1\triangle\partial_i\order{\mathbf{C}}{2}_i = b_1\boldsymbol{\rho}\,, \quad
a_2\triangle\partial_k\order{\mathbf{C}}{2}_k\delta_{ij} + a_3\partial_i\partial_j\partial_k\order{\mathbf{C}}{2}_k + a_4\triangle\partial_i\order{\mathbf{C}}{2}_j = b_2\boldsymbol{\rho}\delta_{ij} + b_3\partial_i\partial_j\mathbf{U}\,,
\end{equation}
where the coefficients \(a_1, \ldots, b_3\) depend on the particular gravity theory under consideration. Hence, they can be solved by an ansatz of the form
\begin{equation}\label{eq:nonmetppn2}
\order{\mathbf{C}}{2}_i = k_1\partial_i\boldsymbol{\chi}\,,
\end{equation}
where the unknown constant \(k_1\) is to be determined by solving the field equations~\eqref{eq:ppnord2}, alongside the PPN parameter \(\gamma\), which is implicitly contained in the constants \(b_1, \ldots, b_3\) via the ansatz~\eqref{eq:metricppn}. The solution is then propagated into the third order field equations, where the nonmetricity terms to be taken into account are the vectors
\begin{equation}
\partial_0\order{\mathbf{Q}}{2}_{i00}\,, \;
\partial_0\order{\mathbf{Q}}{2}_{ijj}\,, \;
\partial_0\order{\mathbf{Q}}{2}_{jij}\,, \;
\partial_i\order{\mathbf{Q}}{3}_{000}\,, \;
\partial_i\order{\mathbf{Q}}{3}_{0jj}\,, \;
\partial_j\order{\mathbf{Q}}{3}_{0ij}\,, \;
\partial_i\order{\mathbf{Q}}{3}_{jj0}\,, \;
\partial_j\order{\mathbf{Q}}{3}_{ij0}\,, \;
\partial_j\order{\mathbf{Q}}{3}_{ji0}\,.
\end{equation}
Inserting all metric and energy-momentum terms, as well as the second-order solution~\eqref{eq:nonmetppn2} obtained in the preceding step, one finds that the third-order field equations can be brought into the generic form
\begin{equation}\label{eq:ppnord3}
a_4\triangle\partial_i\order{\mathbf{C}}{3}_0 = b_4\triangle\mathbf{V}_i + b_5\triangle\mathbf{W}_i\,,
\end{equation}
where the coefficients \(a_4, b_4, b_5\) are again determined by the particular gravity theory whose equations are to be solved. One thus finds that a solution can be obtained from the ansatz
\begin{equation}\label{eq:nonmetppn3}
\order{\mathbf{C}}{3}_0 = k_2\partial_0\boldsymbol{\chi}\,.
\end{equation}
The third-order field equations~\eqref{eq:ppnord3} can be solved by realizing that their left hand side is a pure divergence, while the right hand side, using the definitions~\eqref{eq:ppnpotdef}, splits in the form
\begin{equation}
b_4\triangle\mathbf{V}_i + b_5\triangle\mathbf{W}_i
= \frac{b_4 + b_5}{2}(\mathbf{V}_i + \mathbf{W}_i) - (b_4 - b_5)\mathbf{U}_{,0i}\,,
\end{equation}
where the last term is also a pure divergence, while \(\mathbf{V}_i + \mathbf{W}_i\) is divergence-free as a consequence of the post-Newtonian conservation of energy-momentum. Hence, these two components decouple, and the independent coefficient equations
\begin{equation}
b_4 + b_5 = 0\,, \quad
2a_4k_2 = b_4 - b_5
\end{equation}
determine the PPN parameter \(\alpha_1\), which is contained in the coefficients \(b_4 + b_5\) of the first equation through the metric~\eqref{eq:metricppn}, as well as the coefficient \(k_2\). A similar procedure can then be used to derive and solve the fourth-order field equations. We will not display this general procedure here, since a significantly larger number of terms which are quadratic in second-order perturbations must be included, which would exceed the scope of this article.


\section{Conclusion}
We have presented an extension of the gauge-invariant parametrized post-Newtonian formalism~\cite{Hohmann:2019qgo} to symmetric teleparallel gravity theories, which employ a flat, torsion-free connection next to the metric as a fundamental field variable mediating the gravitational interaction. For this purpose, we have studied the behavior of the connection under gauge transformations, and devised a decomposition into gauge-invariant perturbation variables. Further, we have shown how the perturbations at the second and third velocity orders can be expressed in terms of well-known PPN potentials. Deriving similar expressions for the fourth velocity order is straightforward. The formulas obtained from these calculations can then immediately be applied in order to determine the post-Newtonian limit of various modified symmetric teleparallel gravity theories~\cite{BeltranJimenez:2017tkd,BeltranJimenez:2018vdo,Jarv:2018bgs,Runkla:2018xrv,Hohmann:2018xnb,Koivisto:2019jra}. Also generalizations to other gravity theories involving connections as fundamental dynamical fields may be considered.


\section*{Funding}
The author gratefully acknowledges the full support by the Estonian Research Council through the Personal Research Funding project PRG356, as well as the European Regional Development Fund through the Center of Excellence TK133 ``The Dark Side of the Universe''.

\clearpage

\bibliography{nmginvppn}
\end{document}